\documentstyle[prd,tighten,aps]{revtex}
\begin{document}
\draft
\title{Is the Casimir effect  relevant to sonoluminescence?}
\author{ V.~V.~Nesterenko\thanks{Electronic address:
nestr@thsun1.jinr.dubna.su} and I.~G.~Pirozhenko\thanks{Electronic
address: pirozhen@thsun1.jinr.dubna.su}}
\address{Bogoliubov Laboratory of Theoretical Physics, Joint
Institute for
Nuclear Research \\  141980 Dubna Russia}
\date{\today}
\maketitle
\begin{abstract}
The Casimir energy of a solid ball (or cavity in  an infinite medium)
is calculated by a direct frequency summation using the contour
integration. The dispersion is taken into account, and
the divergences are removed by making use of    the  zeta function
technique. The Casimir energy  of a dielectric ball (or cavity)
turns out to be positive, it being increased when
the radius of the ball decreases. The latter eliminates completely
the possibility of explaining, via the Casimir effect, the
sonoluminescence for bubbles in a liquid.
Besides, the Casimir energy of the air bubbles in water proves to be
immensely
smaller than the amount of the energy emitted in a sonoluminescent
flash. The dispersive effect is shown to be inessential for the
final result.
\end{abstract}


     {\bf 1.} Sonoluminescence  being observed during more than
half century~\cite{sono} has not received satisfactory explanation yet.
As known this phenomenon represents the emission of visual light by
spherical bubbles of air or other gas injected in water and subjected
to an intense acoustic wave in such a way that the radius of bubbles
changes periodically. In the last years of his life Schwinger
proposed~\cite{Schwinger} that the bases of sonoluminescence is
formed by the Casimir effect. While changing the size of bubbles the
zero point energy of the vacuum electromagnetic field (the Casimir
energy) of a cavity in a dielectric medium changes too. According to
Schwinger, it is these changes of the electromagnetic energy that are
emitted as a visual light in sonoluminescent flashes. In Schwinger's
calculations the Casimir energy for the configuration in hand proves
to be of the same order as the energy of the photons in an individual
flash ($\sim 10$~MeV). Other authors obtained results both consistent
with Schwinger's calculation~\cite{Carlson} and differing from it in
10 orders~\cite{Milton,BNP}. This disagreement is basically due to
different methods used for removing the divergences in the problem under
consideration.

    In the present note the calculation of the Casimir energy of a
dielectric ball placed in an endless dielectric
medium (or cavity in this medium) is carried out  under following
conditions.  In the first place a realistic description of dielectric
properties of  media is used which takes into account
dispersion~\cite{endnote}.  On the other hand the most simple and
reliable method for removing the divergences, the zeta function
technique, is applied. Till now these conditions were not combined
in studies of the problem in question.

{\bf 2.}  When calculating the Casimir energy we shall use the
mode-by-mode summation of the eigenfrequencies of the vacuum
electromagnetic oscillations by applying the contour integration
in a complex frequency plane~\cite{NP,BNP}. Consider  a ball
material of which is characterized by permittivity
$\varepsilon _1$ and permeability~$\mu_1$. The ball is assumed
to be placed in an infinite medium with permittivity $\varepsilon_2$
and permeability~$\mu_2$. For this configuration  the frequencies of
transverse-electric (TE)
and transverse-magnetic (TM) modes  are determined by the
equations~\cite{Stratton}
\begin{eqnarray}
\Delta^{\mbox{\scriptsize TE}}_l(a\omega)&\equiv&
\sqrt{\varepsilon_1\mu_2}\,\tilde s'_l(k_1a)\tilde e_l(k_2a) -\sqrt
{\varepsilon_2 \mu_1}\, \tilde s_l(k_1 a) \tilde e_l '(k_2 a) =0,
\label{TE}\\
\Delta^{\mbox{\scriptsize TM}}_l(a\omega)&\equiv&
\sqrt{\varepsilon_2\mu_1}\,
\tilde s'_l(k_1a)\tilde
e_l(k_2a) -\sqrt {\varepsilon_1 \mu_2}\,\tilde s_l(k_1 a)
\tilde e_l '(k_2 a) =0,
\label{TM}
\end{eqnarray}
where
$\tilde s_l(x)=\sqrt{\pi x/2}\,J_{l+1/2}(x)$
and $ \tilde e_l(x)=\sqrt{\pi x/2}\,H^{(1)}_{l+1/2}(x)$
are the Riccati-Bessel functions,
 $k_i=\sqrt{\varepsilon_i \mu_i}\,\omega, \; i=1,2$ are the wave
numbers  inside and outside the ball,  respectively;  prime stands for
the differentiation with respect to the argument ($k_1a$ or  $k_2  a$)
of  the Riccati-Bessel functions.

As usual we define the Casimir energy by the formula
\begin{equation}
\label{def}
E=\frac{1}{2}\sum_s(\omega _s-\bar \omega _s),
\label{energy}
\end{equation}
where $\omega_s$ are the roots of Eqs.~(\ref{TE}) and (\ref{TM}) and
$\bar \omega _s$ are the same roots under condition $a \to \infty $.
Here $s$ is a collective index that stands for a complete set of
indices specifying the roots of Eqs.~(\ref{TE}) and (\ref{TM}):
$s=\{l,m,n\}\quad
l=1, 2, \ldots ;\; m= -(l+1), -l, \ldots, l+1, \; n=1,2, \ldots$.
The  roots of Eqs.~(\ref{TE}) and (\ref{TM}) do not depend
on the azimuthal quantum
number $m$. Therefore the corresponding sum gives a
multiplier $(2l+1)$. Further we use the principle of argument
theorem from the complex analysis in order
to present the sum over $n$ in terms of the contour integral.
As a result Eq.~(\ref{def}) can be rewritten as follows:
\begin{equation}
E=\sum_{l=1}^{\infty}E_l, \quad
E_l=\frac{l+1/2}{2\pi i}\oint\limits_C dz\,z\,\frac{d}{dz}\ln
\frac{\Delta_l^{\mbox{\scriptsize TE}}(az)
\Delta^{\mbox{\scriptsize TM}}_l(az)}{\Delta^{\mbox{
\scriptsize TE}}_l(\infty)
\Delta^{\mbox{\scriptsize TM}}_l(\infty)},
\label{Cauchy}
\end{equation}
where the contour $C$ surrounds, counterclockwise,
the roots of the  frequency
equations (\ref{TE}) and (\ref{TM}) in the  right half-plane.
This contour  can be deformed into a segment  $(- i \Lambda,
i\Lambda)$
of the imaginary axis and
a semicircle of radius $\Lambda $ with $\Lambda \to \infty $.
In this limit the contribution of the semicircle into the
integral (\ref{Cauchy}) vanishes
with the result~\cite{BNP}
\begin{eqnarray}
\label{general}
E_l&=&\frac{l+1/2}{\pi a}\int \limits_0^\infty dy \ln \left \{
\frac{4 e^{-2(q_1-q_2)}}{(\sqrt{\varepsilon_1 \mu_2}+
\sqrt{\varepsilon _2\mu_1})^2}\right.\\
&\times&\left [\sqrt{\varepsilon _1\varepsilon _2\mu_1\mu_2}\left (
(s'_l(q_1)e_l(q_2))^2
+(s_l(q_1)e_l'(q_2))^2
\right)\right.
 -
(\varepsilon _1 \mu_2 +\varepsilon _2\mu_1)
s_l(q_1)s'_l(q_2)e_l(q_2)e_l'(q_2)\bigr ]
\biggr\},\nonumber
\end{eqnarray}
where $q_i=\sqrt{\varepsilon _i\mu_i}\,y,\; i=1,2$ and $s_l(z),
\;e_l(z)$
are the modified Riccati-Bessel functions:
$s_l(z)=(\pi z/2)^{1/2}I_\nu (z),\;e_l(z)=(2z/\pi)^{1/2}K_\nu(z)$,
$\nu=l+1/2$.

Further we will content ourselves by examining the case when both
the media are
nonmagnetic $\mu_1=\mu_2 =1$ and their permittivities $\varepsilon_1,\;
\varepsilon_2$ differ slightly.
In view of this we can put in Eq.~(\ref{general}) $q_1=q_2$
keeping in remain $\varepsilon _1$ and $\varepsilon _2$ exactly.
It  gives
\begin{eqnarray}
E_l&=&\frac{l+1/2}{\pi a}\int_{0}^{\infty}dy \ln \left \{
1-\xi ^2\left [(s_l(y) e_l(y))'\right ]^2
\right \},\label{final}\\
\xi^2&=&\left (\frac{\sqrt{\varepsilon_1}-\sqrt{\varepsilon_2}}{\sqrt
{\varepsilon_1}+\sqrt{\varepsilon_2}}
\right)^2.
\nonumber
\end{eqnarray}
Now we are going to take into account the effect of dispersion
considering the  parameter $\xi^2$ in Eq.~(\ref{final})  as
a function of $y=a \omega/i$. Justification of the mode-by-mode
summation method in applying to dispersive
and absorptive media has been considered in~\cite{Ginzburg}.
For definiteness we put $\varepsilon_1=1+\delta,\; \varepsilon_2
=1,\; \delta \ll 1$, then  $\xi^2\simeq \delta ^2/16$.
We substitute  $\delta$ by
\begin{equation}
\label{delta}
\delta (y) ={\delta_0 }/[1+(y/\nu y_0)^2],\quad \nu
= l+1/2,
\label{dispersion}
\end{equation}
where $\delta _0 $ is a static value of $\delta (y)$ and the parameter
$y_0$ is determined by a "plasma" frequency $\omega_0$:
$y_0 = a \,\omega _0$. The function describing dispersion
in Eq.~(\ref{dispersion}) is a standard one
[one-absorption-frequency Sellmeir dispersion relation]
except for its dependence on $l$. We have introduced this dependence
in order to be able to use  the zeta function technique below.
This complication does not contradict the main goal pursued by using
this function, namely, it should simulate crudely the behaviour of
$\delta (y)$ at large $y$. As known~\cite{LL},
the general theoretical principles lead to the following
properties of the function $\varepsilon (\omega )$ in the
upper half-plane~$\omega $. On the imaginary axis $\omega =i y,\; y>0$
the function $\varepsilon (iy)$
acquires real values, and with increasing $y$ it steadily decreases
from the static value $1+\delta _0>0$ (for dielectrics) to~1.
Obviously formula (\ref{dispersion}) meets these requirements.

Substituting (\ref{delta}) into (\ref{final}) and making use of the
uniform asymptotic expansion for the modified Bessel
functions~\cite{AS} when $l\to \infty$
one obtains
\begin{eqnarray}
E_l\mathop{\simeq}_{l\to \infty}&-&\frac{3}{64 a}
\left (\frac{\delta_0}{4}\right )^2 f_1(a \omega_0)+
\label{asym}
\\
&+&\frac{9}{2^{14}\nu^2}\left(\frac{\delta_0}{4}\right)^2
\left[6 f_2(a \omega_0) -7 f_3(a \omega_0)
\left(\frac{\delta_0}{4}\right)^2
\right]+
{\cal O} (\nu^{-4}),\nonumber
\end{eqnarray}
where
\begin{eqnarray}
f_1(z)&=&\frac{z}{(1+z)^4}\left (
z^3+4\, z^2 +\frac{16}{3}z +\frac{4}{3}\right ){,}\\
f_2(z)&=&\frac{z^3}{(1+z)^7}\left(\frac{521}{9}+\frac{1127}{27} z
+\frac{593}{27} z^2+7 z^3+z^4 \right),\\
f_3(z)&=&\frac{z}{(1+z)^9}\left(\frac{80}{63}+\frac{80}{7}z+
\frac{928}{21}z^2+\frac{1952}{21}z^3+\frac{5960}{63} z^4\right.
\nonumber\\
&&+\left.80 z^5
+\frac{320}{9} z^6 +9 z^7+z^8\right).
\end{eqnarray}

We carry out the summation of the partial Casimir
energies (\ref{final})  with the help
of the zeta function technique~\cite{Od} taking into account
asymptotics~(\ref{asym})
\begin{eqnarray}
\label{renorm}
E=\sum_{l=1}^{\infty}E_l&=&\sum_{l=1}^{\infty}\left [
E_l+\frac{3}{64a}\left (\frac{\delta_0}{4}\right ) ^2f_1(a\omega_0)
-\frac{3}{64 a}\left (\frac{\delta_0}{4}
\right ) ^2f_1(a \omega_0)\right ]  \nonumber \\
&=& \sum_{l=1}^{\infty} \bar E_l - \frac{3}{64 a}
\left (\frac{\delta_0}{4}\right ) ^2 f_1(a \omega_0)
\sum_{l=1}^{\infty}(l+1/2)^0 \\
&=&\sum_{l=1}^{\infty}\bar E_l -\frac{3}{64 a}
\left (\frac{\delta_0}{4}\right )^2
f_1(a \omega _0)[\zeta (0,1/2)-1].
\nonumber
\end{eqnarray}
Here
$\bar E_l=E_l+(3/64a)\left ({\delta_0}/{4}\right )
^2f_1(a\omega _0)$ is the renormalized partial Casimir energy,
$\zeta (s,q)$ is the Hurwitz zeta function. As  $\zeta (0,1/2)=0$,
we get for the Casimir energy~(\ref{renorm})
\begin{equation}
\label{twoterms}
E=\sum_{l=1}^{\infty}\bar E_l +\frac{3}{64 a}
\left (\frac{\delta_0}{4}\right )^2
f_1(a\omega_0).
\end{equation}
With allowance for (\ref{asym}) one can obtain the
estimation for the sum $\sum_{i=1}^{\infty}\bar E_l$
\begin{eqnarray}
\sum_{i=1}^{\infty}\bar E_l&\simeq&
\frac{9}{2^{14}}\left(\frac{\delta_0}{4}\right)^2
\left[6 f_2(a \omega_0) -7 f_3(a \omega_0)
\left(\frac{\delta_0}{4}\right)^2
\right]\sum_{l=1}^{\infty}\frac{1}{(l+1/2)^2}\nonumber\\
&=&
\frac{9}{2^{14}}\left(\frac{\delta_0}{4}\right)^2
\left[6 f_2(a \omega_0) -7 f_3(a \omega_0)
\left(\frac{\delta_0}{4}\right)^2
\right]\left(\frac{\pi^2}{2}-4\right)\nonumber\\
&=&
5.135\times10^{-4}\left(\frac{\delta_0}{4}\right)^2
\left[6 f_2(a \omega_0) -7 f_3(a \omega_0)
\left(\frac{\delta_0}{4}\right)^2
\right].
\end{eqnarray}
Thus the Casimir energy of a
dielectric ball is
\begin{equation}
\label{final2}
E\simeq\frac{3}{64 a} \left (\frac{\delta_0}{4}\right )^2
\left\{f_1 \,(a\omega_0)+0.066 \, f_2(a\omega_0)-
0.0048\, \delta_0^2 \,f_3(a\omega_0)\right\},
\end{equation}
dispersion resulting only in the positive functions
$f_i(a\omega_0),\; i=1,2,3.$
When  $z$ increases the functions $f_i(z)$  approach~1 (see Fig.~1),
and (\ref{final2}) turns into the expression for the partial Casimir
energy of a solid ball without dispersion~\cite{BNP}.

Considering the behaviour of the functions $f_i(z)$ (see Fig. 1)
one concludes that the main contribution, with a few percents
accuracy, gives the first term in braces in Eq. (\ref{final2})
with the result
\begin{equation}
\label{final3}
E\simeq\frac{3}{64 a} \left (\frac{\delta_0}{4}\right )^2
f_1(a\omega_0).
\end{equation}
Obviously the change of the energy sign or a considerable
increasing its magnitude due to the dispersion
effect~\cite{Brevik}   is out of the question.

Let us estimate the value of $f_1(a \omega_0)$.
The parameter $\omega _0$ can be determined by demanding
that at this frequency the photons do not `feel' the interface
between two media. This condition will be
certainly met when the wave length of photon is less
than the interatomic distance
in media $d \sim 10^{-8}$~cm.
Actually it is the condition of applicability of the
macroscopic description
of  dielectric media~\cite{LL}.
Sonoluminescence is observed with the air bubbles in
water~\cite{sono},
the radius of bubbles being $a \sim 10^{-4}$~cm.
Hence it follows that $a\,\omega_0 \simeq a/d =10^4$
and $f_1(10^4)=0.999\ldots $.
Thus the allowance for the dispersion in calculating
the Casimir energy of a dielectric
ball (or spherical cavity in a slab of a dielectric)
practically has no effect on the final result.

Certainly the real picture of dispersion in the whole frequency range
$0<\omega <\infty $  for any dielectric, including water, is
exceedingly complicated and cannot be described by a simple
equation  (\ref{delta})  with a single
parameter~$\omega_0$. As known absorption of the electromagnetic waves
in water and, as
a consequence, their dispersion take place already in
the radio frequency band. Putting
in this case $\lambda \sim 10^4$~cm, we obtain
$a\,\omega _0\sim 1$  and $f_1(1)=0.729\ldots$.
From here  one can infer that the effective value of  $a\,\omega_0$
should be less than~$10^4$.
In order for a more precise evaluation of this parameter to be done
a more detailed consideration of the dispersion mechanism is needed.
Obviously this may lead only to diminution of the absolute value
of the Casimir energy. However
this issue is beyond the scope of the present paper for the main
conclusion (see below)
does not depend on this point.

It is worth noting  two peculiarities of the final formula
(\ref{final3}).
When the radius of the bubble decreases its Casimir energy
increases. This behaviour
is completely opposite to one needed for  explanation of
sonoluminescence  (as known, emission of light
takes place at the end of collapsing the bubbles in liquid).
Besides, this energy is immensely smaller than the amount of energy
emitted in a separate sonoluminescent flash ($\sim 10$~MeV).
Actually taking  $a=10^{-4}$~cm and $\delta_0=3/4$
(water) we arrive at  $E\simeq 5\cdot 10^{-3}$~eV.

Thus the results of this paper unambiguously testify that the
Casimir effect is irrelevant to sonoluminescence.

This work was accomplished with financial support of Russian
Foundation of Fundamental Research (Grant ü 97-01-00745).



\begin{thebibliography}{99}
\bibitem{endnote} There is a point of view~\cite{Candelas,Brevik}
that the dispersion effects may essentially affect the final result
when calculating the Casimir energy.
\end{thebibliography}
\end{document}